# Spectroscopic study of olivine-bearing rocks and its relevance to the ExoMars rover mission

Marco Veneranda[1], Jose Antonio Manrique[1], Guillermo Lopez-Reyes[1], Jesús Medina[1], Imanol Torre-Fdez[2], Kepa Castro[2], Juan Manuel Madariaga[2], Cateline Lanz[3], Francois Poulet[3], Agata M. Krzesińska[4], Helge Hellevang[4], Stephanie C. Werner[4], Fernando Rull[1]

[1] *Department of Condensed Matter Physics, Crystallography and Mineralogy, Univ. of Valladolid, Spain. Ave. Francisco Vallés, 8, Boecillo, 47151 Spain.* marco.veneranda.87@gmail.com
[2] *Department of Analytical Chemistry, University of the Basque Country (UPV/EHU), 48080 Bilbao, Spain*
[3] *Institut d'Astrophysique Spatiale, CNRS/ Université Paris-Sud, France*
[4] *Department of Geosciences,* Centre for Earth Evolution and Dynamics, *University of Oslo, Norway*

**Abstract**

We present the compositional analysis of three terrestrial analogues of Martian olivine-bearing rocks derived from both laboratory and flight-derived analytical instruments. In the first step, state-of-the-art spectroscopic (XRF, NIR and Raman) and diffractometric (XRD) laboratory systems were complementary used. Besides providing a detailed mineralogical and geochemical characterization of the samples, results comparison shed light on the advantages ensured by the combined use of Raman and NIR techniques, being these the spectroscopic instruments that will soon deploy (2021) on Mars as part of the ExoMars/ESA rover payload. In order to extrapolate valuable indicators of the mineralogical data that could derive from the ExoMars/Raman Laser Spectrometer (RLS), laboratory results were then compared with the molecular data gathered through the RLS ExoMars Simulator. Beside correctly identifying all major phases (feldspar, pyroxene and olivine), the RLS ExoMars Simulator confirmed the presence of additional minor compounds (i.e. hematite and apatite) that were not detected by complementary techniques. Furthermore, concerning the in-depth study of olivine grains, the RLS ExoMars simulator was able to effectively detect the shifting of the characteristic double peak around 820 and 850 $cm^{-1}$, from which the Fe-Mg content of the analysed crystals can be extrapolated. Considering that olivine is one of the main mineral phases of the ExoMars landing site (Oxia Planum), this study suggests that the ExoMars/RLS system has the potential to provide detailed information about the elemental composition of olivine on Mars.

**1 Introduction**

The current knowledge concerning the mineralogical composition of Mars has been achieved through the interpretation of global data sets remotely recorded by Martian orbiters (i.e. Mars Express [1,2] and Mars Reconnaissance Orbiter [3-6]) and in-situ analyses performed by landers (i.e. Viking 1 and 2 [7], Pathfinder [8] and Phoenix [9]) and rovers (i.e. Spirit [10], Opportunity [10,11] and Curiosity [12,13]). These data, combined with the additional information gathered on Earth through the in-depth study of Martian meteorite fragments [14-16] and terrestrial analogue materials [17,18], help to shed light on the geological and environmental evolution of Mars.

In this framework, the aim of the Planetary Terrestrial Analogues Library (PTAL) project [19] is to provide the scientific community with a multi-instrument spectral database of natural, experimental and artificial terrestrial analogue materials resembling the geochemical and mineralogical composition of Martian rocks and soils. As detailed in the work of M. Veneranda et al. (2019) [20], the PTAL database presents over 90 natural analogue samples that can be divided in two main categories. On one hand, unaltered geological samples congruent to Martian precursor materials, including several mafic and ultramafic lithologies. On the other hand, altered materials covering a broad range of aqueous alteration processes underwent in both hydrothermal and surface weathering conditions. Better understanding of these two aspects - mafic composition and how it influences aqueous alteration - is important for the correct interpretation of Martian data. As such, detailed characterization of the PTAL analogues is a valuable asset to support the interpretation of the spectroscopic data that will be gathered by the ExoMars rover analytical suite.

Led by the European Space Agency (ESA), the ExoMars mission will deliver to Mars an exploration rover equipped with a set of panoramic and contact instruments (Pasteur payload) to study geological samples collected from the Martian subsurface down to a depth of 2 meters [21]. The sample preparation and distribution system (SPDS) of the Pasteur payload will crush the samples to be delivered to the MicrOmega Near Infrared Spectrometer [22], the RLS Raman Laser Spectrometer (RLS) [23] and the MOMA Laser Desorption Mass Spectrometer [24]. The aim is to identify the mineralogical composition of the samples and to detect the possible presence of organic compounds on the same spot of the same sample, this being possible for the first time in an in-situ rover mission [25]. Combined RLS and MicrOmega data will be of critical importance in the selection of the materials to be analyzed by the Mars Organic Molecule Analyzer (MOMA), whose goal is to extract, separate and identify the organic molecules potentially preserved in the geological samples [26]. Considering the key role played by spectroscopic tools in the discrimination of the geological samples to be analyzed by MOMA, it is important to evaluate their versatility in the characterization of the Martian's mineral diversity through the study of terrestrial analogues.

In response to this need, the present work summarizes the multi-analytical characterization of rock samples from Iceland (Reykjanes), being those widely recognized as optimal terrestrial analogues of Martian olivine-bearing rocks [27-29]. Indeed, as presented by A. Ody et al. (2013) [30], the spectroscopic data collected by the OMEGA instrument on board the Mars Express/ESA orbiter identified five major geological Martian settings presenting high concentration of olivine. The remote detection of this mineral was confirmed in several locations by in-situ measurements performed through exploration rovers [31-33] as well as by multi-analytical studies carried out on Earth on Martian meteorites fragments [34,35]. In the framework of the ESA/ExoMars rover mission, the study of olivine-bearing rocks acquires a critical importance since, according to remote CRISM data, olivine is one of the major mineral phases detected on the landing site (Oxia Planum) [36].Thus, the analytical study of olivine-bearing analogues (i.e picritic basalts) has a critical scientific relevance for several reasons. First, being the result of the crystallization of mantle-derived magma, its study provides the opportunity to deepen the knowledge about the composition of the planet's mantle [37]. Second, from the study of olivine-degradation products it is possible to extrapolate important inferences about the environmental conditions triggering their alteration [38]. Third, considering that under hydrated conditions the degradation of olivine leads to the formation of phyllosilicates such as smectites and serpentines [39], altered picritic basalts are considered optimal targets to search for biomarkers on Mars [40].

In light of the scientific relevance of studying olivine-bearing analogues to support the forthcoming ExoMars mission, the Icelandic samples selected for this study were analyzed by using a combination of laboratory and flight-derived analytical instruments. At first, spectroscopic and diffractometric laboratory systems were used to obtain the detailed mineralogical and geochemical characterization of the analogues. Afterwards, laboratory results were compared with those obtained from the RLS ExoMars Simulator [41], a Raman system developed by the UVa-CSIC-CAB Associated Unit ERICA (Spain) to effectively predict the scientific capabilities of RLS/ExoMars instrument that will be soon deployed on Mars.

On a general perspective, the present work aims at providing the whole set of spectroscopic data to be included in the PTAL analogue database [20]. Furthermore, by comparing the results obtained from the RLS ExoMars Simulator with those provided by laboratory spectroscopic and diffractometric systems, it seeks to provide a valuable indicator of the role that the ExoMars/RLS system could play in the proper characterization of olivine-bearing rocks and soils on Mars.

## 2 Materials and methods

### 2.1 Analyzed materials

The rocks in Reykjanes represent basaltic lavas from fissure-fed shield volcanos and sourced from Atlantic mid-oceanic ridge [42]. As the source of these lavas is depleted mantle source, the rocks represent basaltic, tholeiitic and picritic lithologies. However, lavas may embed xenoliths that are gabbroic in composition. PTAL geologists scoured the Reykjanes region to identify the optimal analogue samples to be included in the database. According to the information gathered during the field research, three rock samples containing considerable and various amount of olivine phenocrysts (($Mg,Fe$)$_2SiO_4$) were collected from areas were original rocks visually presented weak or null alteration features. In detail, the sampling site of IS16-0001 (355.8 g) and IS16-0002 (472.8 g) analogues is located at the geographic coordinates N63 49 01.7 W22 39 03.1, while sample IS16-0013 (318.2 g) was collected at N63 48 58.3 W22 39 38.8.

Prior to spectroscopic and diffractometric studies, thin sections were prepared to reveal the textural details of rocks. To do so, blue stained epoxy resin was used to glue rock fragments to glass microscope slides. Then, thin sections were prepared by polishing the glued samples to a final thickness of 30 μm. According to polarized light microscopy analyses, texture and mineralogy observed on samples IS16-0001 and IS16-0002 is typical for tholeiitic basalts (Figure SM1). On one hand, sample IS16-0001 contains abundant, large subhedral-euhedral phenocrysts crystals (olivine and plagioclase) set in fine-grained matrix that itself is a mixture of Ca-rich pyroxene and plagioclase. On the other hand, sample IS16-0002 contains lower amount of olivine phenocrysts than sample IS16-0001. Olivine crystals are embedded in a fine-grained pyroxene-plagioclase matrix. Sample IS16-0013 contains minor amount of olivine, that occurs as small anhedral crystals only. Lath-shaped plagioclase crystals are commonly dispersed throughout the sample. Millimetre-sized intergrowths of pyroxene with plagioclase are distinct in this sample. Matrix is fine-grained and composed of plagioclase and pyroxene. Texturally and mineralogically, this rock resembles basaltic-gabbroic lithology rather than tholeiite (Figure SM1).

After preliminary thin section observations, further samples fragments were used to prepare coarse and fine powders. Coarse powders (grains size up to 500 µm) were obtained by crushing rock fragments in an agate mill (McCrone Micronizer Mill) for 12 minutes to reproduce the granulometry of the crushed materials that the ExoMars/SPDS will deliver to RLS and MicrOmega instruments on board the rover. A fraction of the coarse powder was then further milled to obtain the grain size necessary to perform optimal XRD analysis (below 65 µm).

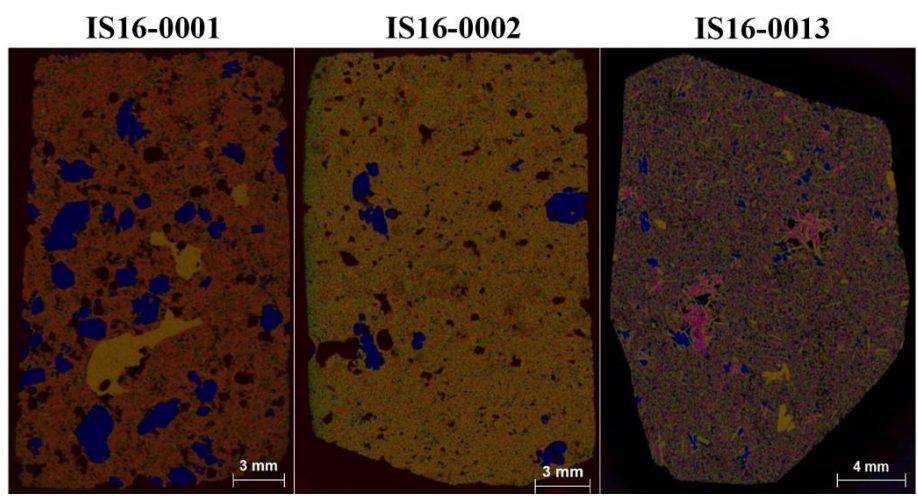

*Figure SM1: on the left X-ray elemental composite maps showing mineral composition of studied analogues. Red-Ca, Green-Al, Blue-Mg. In this color code, olivine appears blue, plagioclase appears yellow. Pyroxene has varying colors, from red to purple, reflecting relative content of Mg and Ca, that is: diopside (Ca-px) is red, augite (Ca>Mg-px) is light purple and pigeonite (Ca~Mg-px) is dark purple.*

## 2.2 Analytical instruments

In the first step, a comprehensive characterization of the selected samples was performed by combining the use of spectroscopic and diffractometric laboratory systems. The obtained results were then used as a reference in the evaluation of the scientific capabilities of the RLS ExoMars Simulator.

### 2.2.1 Laboratory instruments

*Raman Spectroscopy*

For the molecular characterization of powdered Icelandic basalts, a Renishaw inVia Raman micro-spectrometer (Renishaw) equipped with a continuous 532 nm excitation diode laser was employed. The instrument is equipped with a charge-coupled device detector (Peltier cooled) and is coupled to a confocal microscope with interchangeable long WD objective of 10x, 20x, 50x and 100x (used in this work). Between 55 and 65 spectra were collected from each powdered sample by focusing the excitation laser on the most interesting mineral grains(spot size of ≈20μm, working distance =3.4 mm), which were visually selected by the operator through the camera coupled to the Raman system.

Raman spectra (mean spectral resolution of 1 cm$^{-1}$) were then acquired in a range of 50–3800 cm$^{-1}$ for 5–60 s and, while a number of scans between 5 and 20 were accumulated to improve the signal-to-noise ratio. During analysis, the excitation source was set at medium to low power (not more than 10 mW on the sample) to avoid the thermal decomposition of the materials under study. Data were acquired using the Wire 4.2 software package (Renishaw).

*Near Infrared Spectroscopy (NIR)*

Complementary to Raman data, further molecular analysis was done using a reflectance spectrometer in the near-infrared (0.8–4.2 μm). For this purpose, a Fourier Transform spectrometer (PerkinElmer Spectrum 100N FTNIR) was employed under ambient temperature and pressure conditions. For the study of powdered samples, the spectral resolution and the spectral sampling were set to 4 cm$^{-1}$ and 0.35 nm, respectively. The collecting spot size of about 1 mm ensured a representative bulk characterization of all constituents of the powdered material. To calibrate the sample reflectance spectrum, reference spectra were acquired using an Infragold and a Spectralon 99% (Labsphere). An automated correcting mode on the instrument was used to better correct the OH signatures due to ambient air (slightly noisy data are spotted around 1.4, 1.9, and 2.7 μm). Compared to the NIR instrument onboard the ExoMars rover, the spectrometer used in this work is appropriate to acquire single spectra (spot of 500 μm) of powdered samples where all initial rock constituents are mixed, thus emphasizing the averaged composition of the target. On the other hand, MicrOmega is dedicated to observe heterogeneous rocks where inclusions can be detected at the pixel size of the detector, obtaining monochromatic images with a high resolution of 20 μm × 20 μm.

*X-ray fluorescence (XRF)*

The elemental composition of Icelandic basalts was determined through the M4-300+ TORNADO Energy Dispersive X-ray fluorescence spectrometer (ED-XRF, Bruker Nano GmbH), equipped with a Rh X-ray tube, polycapillar lenses and a XFlash® silicon drift detector. On one side, point by point analyses of very fine-grained powders were performed to obtain a quantitative estimation of the elemental composition of minerals in the sample. To improve the detection limit for light elements, measurements were performed under vacuum (20 mbar). Data were acquired using the M4-300+ TORNADO software (Bruker Nano GmbH).

*X-ray diffractometry (XRD)*

X-ray diffraction patterns were obtained using a D8 Advance diffractometer (Bruker) equipped with a Cu X-ray tube (wavelength 1.54 Å) and a LynxEye detector. The random powder mounts of the bulk samples were analyzed in a step scan mode from 2 to 65° 2θ with a step increment in 2θ of 0.01 and a count time of 0.3 seconds per step. The mineral phases were identified with a DIFFRAC.EVA search-match module coupled with Crystallography open database and PDF-2. Afterwards, mineral phases were quantified with a Rietveld refinement routine using the PROFEX-BGMN-Bundle software [43].

*2.2.2 Flight-derived analytical instruments*

*RLS ExoMars Simulator*

The RLS ExoMars Simulator is considered the most reliable tool to effectively simulate the analytical procedure followed by the Raman system instrument onboard of the ExoMars/ESA rover [44]. The instrument setup includes a continuous 532 nm excitation laser (BWN-532, B&WTek), a BTC162 high resolution TE Cooled CCD Array spectrometer (B&WTek) and an optical head with a long WD objective of 50x. Unlike the InVIa spectrometer, the spot of analysis (≈ 50 µm) and the working distance (≈ 15 mm) of this instrument are closely resembling those of the RLS flight model. The RLS ExoMars Simulator is coupled to a replicate of the Sample Preparation and Distribution System (SPDS) of the Pasteur suite and is capable of working in automatic mode by using the same algorithms that will be employed by the RLS on Mars. Thus, the instrument is capable of autonomously performing several operations such as sequential analysis, autofocus, optimization of the signal to noise ratio and automatic selection of the best integration time/number of accumulations values [45].

To evaluate the accuracy of the algorithms developed for automatic analysis, powdered samples were analyzed by simulating the operational conditions of the RLS ExoMars system on Mars. For each sample, between 20 and 40 spots were selected by moving the positioners in the x-axis at steps of 150 microns and spectra were collected by automatically adjusting the several parameters described above. All Raman spectra were collected in a range of 70–4200 $cm^{-1}$ with a spectral resolution of 6–10 $cm^{-1}$ (as the RLS flight model) and data acquisition was performed using a custom developed software based on LabVIEW 2013 (National Instruments). Raman spectra were finally treated and interpreted by means of the IDAT/Spectpro software [46].

**3 Results and discussion:**

**3.1 Analysis with laboratory systems**

**3.1.1 X-ray diffractometry of fine powders (D8 Advance)**

According to XRD analysis, plagioclase, forsterite, augite and ankerite dominate the diffractograms obtained from the three analysed samples (Figure 01). By comparison with reference diffraction patterns, the best match for the plagioclase result to be labradorite ((Ca,Na)(Al,Si)4O8), an intermediate member of albite-anorthite solid-solution (An 50-70 mol%; atomic ratio Ca/(Ca + Na) of 0.50-0.70). The intensity of the peaks in the diffractograms suggests a similar plagioclase content in the three samples. Volume of olivine varies for each sample from negligible in sample IS16-0013 to the most prominent peaks in sample IS16-0001 (see Figure 01). In the case of sample IS16-0001, additional traces of hematite ($Fe_2O_3$) and ilmenite (FeTiO) were also detected. Being carbonates, the detection of ankerite ($Ca(Fe,Mg,Mn)(CO_3)_2$) and calcite ($CaCO_3$) in the three diffractograms proves that the three analogues have been altered to some degree.

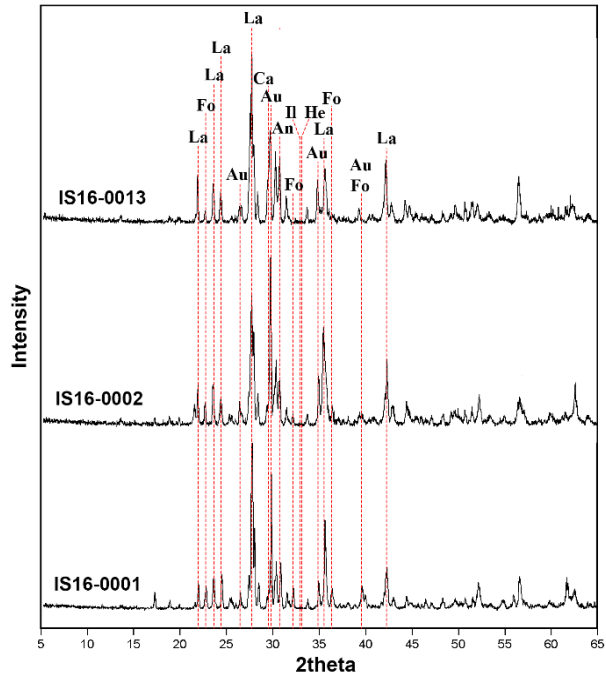

*Figure 01: Comparison of XRD measurements of IS16-0001, IS16-0002, and IS16-0013 samples. Labels: La (labradorite); Fo (forsterite); Au (augite); An (ankerite); Il (ilmenite); Ca (calcite); He (hematite).*

Quantitative phase analysis was performed through the Profex software [47] by using the Rietveld refinement program BGMN [48]. The obtained results indicate that olivine content varies from about 25% in IS16-0001 to 3% in IS16-0013 (see Table 01). The content of pyroxene (augite) changes correspondingly from about 35% for IS16-0001 to more than 50% for IS16-0013. Plagioclase is similar in all samples at about 40%.

### 3.1.2 NIR spectroscopy of coarse powders (Spectrum 100N FTNIR)

NIR spectra from samples IS16-0001, IS16-0002 and IS16-0013 are presented in Figure 02. The spectra of three analogues show strong and broad silicate signatures (centered near 1.0 and 2.25 µm; called bands 1 and 2, respectively). The silicate features can be attributed to both olivine (band 1) and pyroxene (bands 1 and 2 [49]). In addition those, pyroxenes also exhibit a weaker absorption band at 1.2 µm. This absorption is best defined in Ca-saturated pyroxenes [50]. All three pyroxene absorptions are the result of crystal field transitions of iron in octahedral coordination [49].

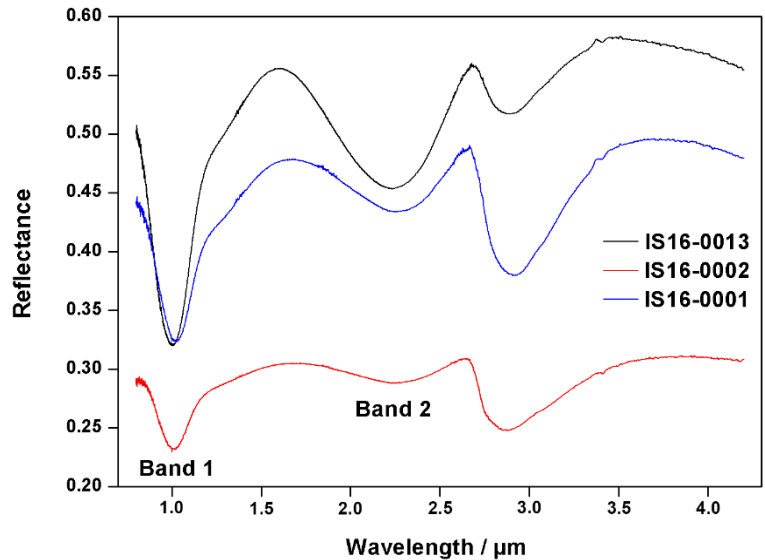

*Figure 02: NIR spectra obtained from sample IS16-0001, IS16-0002 and IS16-0013. Three bands centered near 1, 1.2 and 2 µm can be attributed to pyroxenes.*

The substitution of different sized cations, such as calcium, iron, and magnesium, has significant effects on the position and shape of the pyroxene 1 and 2 µm [49,51]. Of special interest is the variation of the band centers depending of Ca content. The band centers of low-calcium pyroxenes (e.g., orthopyroxenes and low-calcium clinopyroxenes, LCP) occur near 0.9 and 1.8 µm while the band centers of high-calcium pyroxenes (HCP) are located near 1.05 and 2.3 µm. All three samples exhibit band centers at 1.0 µm and ~2.25 µm. This implies the presence of HCP as a dominant pyroxene phase with possibly a small amount of LCP, especially for sample IS16-0013 for which the 2 µm band is centered at smaller wavelength than the other ones. The identification of olivine is based on its broad 1 µm band due to $Fe^{2+}$ ion in distorted octahedral sites. The shape of the band at 1 µm of sample IS16-0001 combined to a lower 2 µm band depth suggests the presence of olivine in this sample.

The broad 2.9 µm band is associated to presence of water. The water band on IS16-0002 spectra has a different position than the two others as a feature appears at 2.75 µm, which is diagnostic of OH and $H_2O$ stretching band in phyllosilicates. The interpretation of this vibrational signature suggests that the silicates in this sample are possibly aqueously altered. However, the lack of overtones and combinations in the 1-2.5 µm range indicates a very low degree of alteration in the analyzed powders. The 2.9 µm band in sample IS16-0013 is very small compared to the band depth of silicates, suggesting a lower alteration degree than IS16-0001 and IS16-0002.

Comparing with XRD results, plagioclase was not detected since they are largely featureless in the IR range. It is nevertheless possible to retrieve abundance of this featureless phase by modeling the spectra [52]. Following the methodology previously developed for Mars surface remote study [53], the three spectra have been reproduced by simulating the Light-Matter Interaction of a mineral mixture from a radiative transfer modeling based on the Shkuratov theory [54]. As the optical constants are not known for all minerals, proxies were used to mimic major mineral families. As discussed in Poulet et al. (2009), HCP is modeled by diopside $((Ca,Mg,Fe)_2(Si,Al)_2O_6)$ and augite, while pigeonite $((Ca,Mg,Fe)(Mg,Fe)Si_2O_6)$ is included in the mixture as Ca-poor pyroxene end-member. The band positions of augite are slightly shifted to shorter wavelengths than diopside by a few tens of nm, which is consistent with the lower Ca content in augite. In the same way, plagioclase is modeled through labradorite with small inclusions of iron oxide ilmenite. Olivine as forsterite end-member is also included in the model. An example of the modeling results is provided in Figure 03.

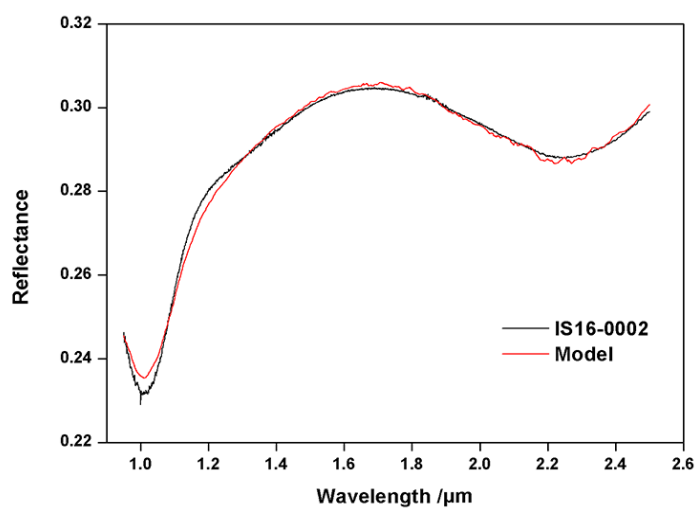

*Figure 03: Example of NIR spectrum and its best-fit model (here sample IS16-0002).*

Table 01 summarizes the modeled abundances for the three samples from the spectral modeling in the 1-2.5 µm range. This reduced spectral range was selected for the modeling because of the lack of optical constants in the other part of the spectral range. Derived abundances of end-members are accurate to within 10% for the most abundant phases [52]. Furthermore, the model can hardly reproduce spectral signature resulting from olivine with small grain size when abundance is below ~10% [55].

As displayed in Table 01, for IS16-0001 the model gives a strong abundance of HCP pyroxene (diopside as proxy, 44%) and plagioclase (labradorite, 44%). Augite (9%) is also required, together with minor amount of olivine (forsterite, >5% with 100 µm grain size), and ilmenite (>1%). For sample IS16-0002, the model suggests HCP (35% of diopside and 33% of augite) and plagioclase (31%) as major mineral phases, while olivine and ilmenite are present in concentrations below 5%. According to the model, sample IS16-0013 presents high amounts of HCP (augite, 78%) and plagioclase (15), together with LCP pigeonite ((Ca,Mg,Fe)(Mg,Fe)$Si_2O_6$ ,6%) and ilmenite (>1%) as minor mineral phases. The presence of pigeonite is required to fit the position of the 2 µm band shifted to a lower wavelength than 2.3 µm.

### 3.1.3 Raman spectroscopy of coarse powders (InVia spectrometer)

In the case of sample IS16-0001, many Raman spectra displayed a double peak around 665 and 1010 cm$^{-1}$ accompanied by secondary signals at 324 and 390 cm$^{-1}$, which are the characteristic vibrational features of minerals belonging to the pyroxene group. As shown in Figure 04, the differences between the obtained profiles suggest the presence of two minerals. Comparing these data with the Raman analysis presented in the work of S. Andó and E. Garzanti (2014) [56], it was established that the spectrum in Figure 04b, presenting an additional peak at 136 cm$^{-1}$, is characteristic of diopside ((Ca,Mg)$_2$(Si,Al)$_2$O$_6$), while the spectrum in Figure 04a (obtained from the analysis of a different grain) fits the augite pattern ((CaMgFe)2Si$_2$O$_6$). This interpretation agrees perfectly with the information provided by XRD and NIR techniques.

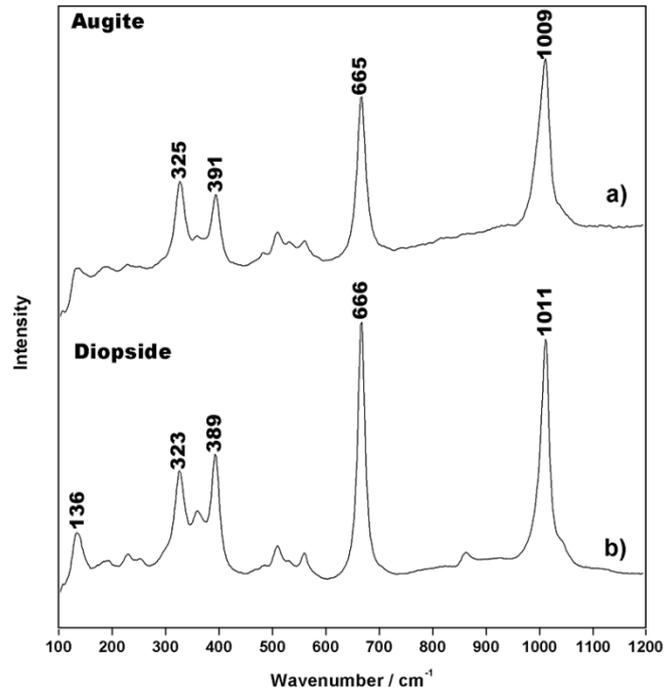

*Figure 04: Raman spectra of augite (a) and diopside (b), collected from sample IS16-0001 by means of the laboratory InVia spectrometer.*

A considerable number of analyses detected a double peak around 482 and 508 cm$^{-1}$, proving the presence of plagioclases. Looking at the position of secondary signals, two minerals can be distinguished. As shown in Figure 05a, the intense peak at 177cm$^{-1}$ is characteristic of labradorite, a plagioclase having a Ca/(Ca + Na) value between 0.50-0.with 50-70 mol% of An [57], while the peaks at 198, 285, 405 and 562 cm$^{-1}$ confirmed the detection of high temperature anorthite (Figure 05b, plagioclase endmember having a Ca/(Ca + Na) value between 0.9-1).

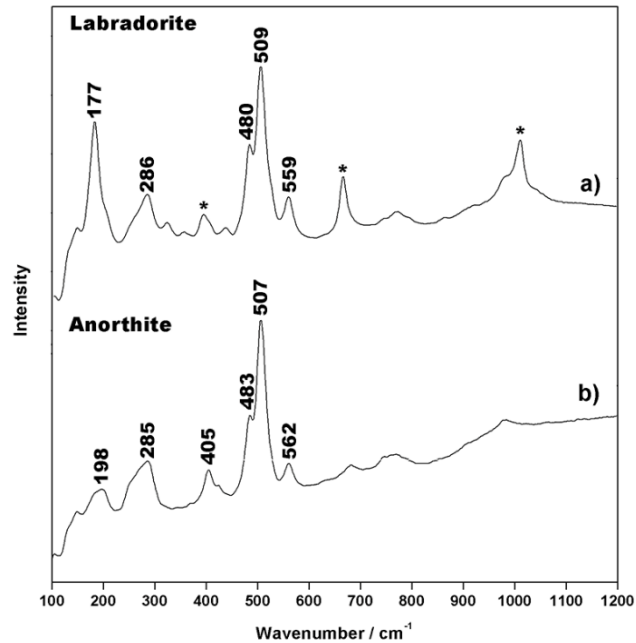

*Figure 05: Raman spectra of labradorite (a) and anorthite (b), collected from sample IS16-0001 by means of the laboratory InVia spectrometer. * Pyroxene Raman signals proceeding from micrometric grains partially covering the analysed labradorite and anorthite crystals.*

As underlined by XRD and NIR analysis, IS16-0001 is the sample having the highest concentration of olivine. By means of Raman analysis, the main peaks of olivine were detected around 820 and 850 cm$^{-1}$, while the secondary signals appeared around 315, 420, 530, 915 and 955 cm$^{-1}$. However, it must be highlighted that, depending on the crystal grain under analysis, the exact position of the two main peaks ranges between 816 and 823 cm$^{-1}$ ($\kappa_2$: Si-O symmetric stretching band A$_g$(Si-O)$_{a-str}$) and 848 and 855 cm$^{-1}$ ($\kappa_1$: Si-O asymmetric stretching band A$_g$(Si-O)$_{s-str}$) respectively (Figure 06).

As detailed in the work of Torre et al. [47], this phenomenon is related to the concentration of iron and magnesium ions in the crystalline structure of olivine, being this a solid solution between fayalite (Fa, Fe$_2$SiO$_4$) and forsterite (Fo, Mg$_2$SiO$_4$) end-members. Considering that the position of the double peak varies depending on the grain under analysis, it was deduced that the sample contains olivine crystals with different Fo/Fa ratio.

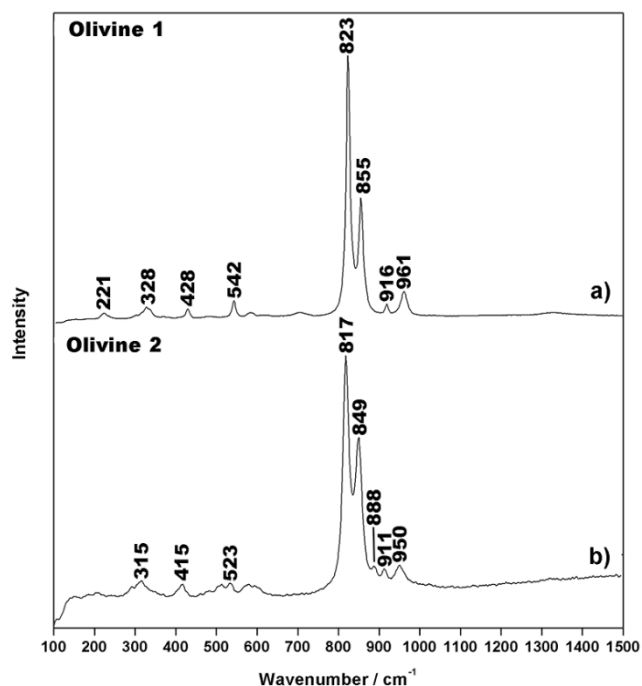

*Figure 06: Raman spectra of different olivine grains from sample IS16-0001, analysed by means of the laboratory InVia spectrometer. The displacement of the main Raman peaks is related to the Fo/Fa ratio of the solid solution.*

Taking into account the linear relationship between the displacement of Raman peaks and the Fo/Fa ratio, calibration lines were proposed to calculate the exact composition of olivine crystals through the refined interpretation of Raman spectra [48,49]. According to the work of Torre et al., the equation presented by Mouri and Enami [48], which is based on the evaluation of the displacement of the $\kappa_1$ band, was identified as the most reliable semi-quantification method (Eq.1).

$$\%Mg = (-610.65 + 1.3981\kappa_1 - 0.00079869\kappa_1^2)*100 \; [Eq. 1]$$

According to this equation, the average composition of olivine crystals in sample IS16-0001 varies between $Fo_{86}Fa_{14}$ and $Fo_{64}Fa_{36}$.

In addition to pyroxene, feldspar and olivine, the detection of a limited number of Raman spectra with an intense band at 679 cm$^{-1}$ confirmed the presence of ilmenite ($FeTiO_2$) as a minor phase. Finally, two spectra of hematite, recognizable by the characteristic peaks at 225, 294, 409, 500, 609, 664 and 1318 cm$^{-1}$ were also collected. Concerning the detection of minor alteration products, hematite, calcite (main Raman peak at 1086cm$^{-1}$) and carbon (Raman bands at 1350, and 1585 cm$^{-1}$) were only detected in few spots. This data agrees with the information derived from complementary techniques and thin section observations, demonstrating the low degree of alteration of this material.

Most of the Raman spectra obtained from sample IS16-0002 returned the vibrational profiles of pyroxenes and plagioclase. Raman analysis also detected a considerable amount of olivine and hematite grains (six and five, respectively), while ilmenite and carbon were found in a very few points of interest (less than three). According to spectra comparison, labradorite was the only plagioclase detected in this analogue, while both diopside and augite were present. With regards to olivine, the position of the Si-O symmetric stretching peak constantly appeared at 822 cm$^{-1}$, while the Si-O asymmetric stretching peak was always found at 854 cm$^{-1}$. By employing the previous equation, a solid solution of $Fo_{83}Fa_{17}$ was estimated. Concerning weathering products, the percentage of points of interest in which hematite was detected was below 10%, indicating a low degree of alteration (Figure 07).

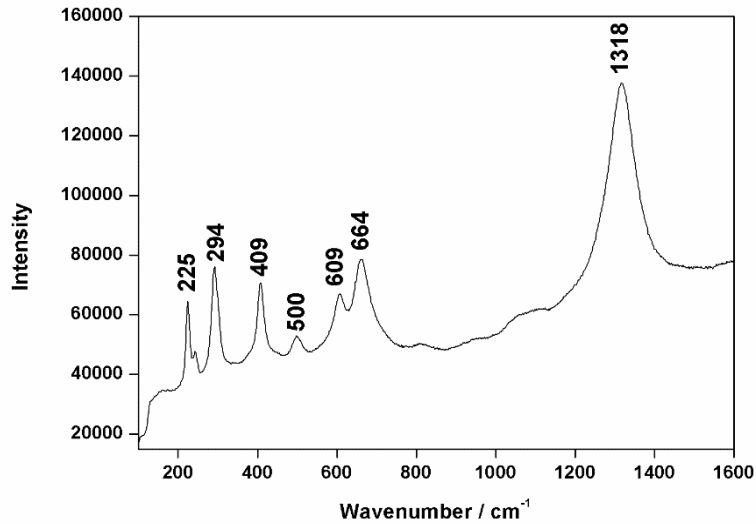

*Figure 07: Raman spectrum of hematite, collected from sample IS16-0002 by means of the laboratory InVia spectrometer.*

From the Raman analysis of sample IS16-0013 pyroxene spectra were mostly obtained. Again, the vibrational differences among spectra clearly proved the co-presence of diopside and augite. However, in contrast to samples IS16-0001 and IS16-0002, a third pyroxene was identified. Indeed, as shown in Figure 08, some spectra were characterized by the displacement of the two main peaks (from 666 to 673 cm$^{-1}$ and from 1011 to 1001 cm$^{-1}$ respectively), the appearance of a peak shoulder at 657cm$^{-1}$ and the substitution of the two secondary bands at 323 and 390 cm$^{-1}$ with a medium intensity peak at 330 cm$^{-1}$. By comparing the spectrum with a spectra database, it was deduced that this profile could be assigned to both pigeonite ($(Mg,Fe\ Ca,)_2Si_2O_6$) and hypersthene ($(Mg,Fe)_2Si_2O_6$) mineral. Considering the information provided by NIR data, this spectrum was therefore assigned to pigeonite.

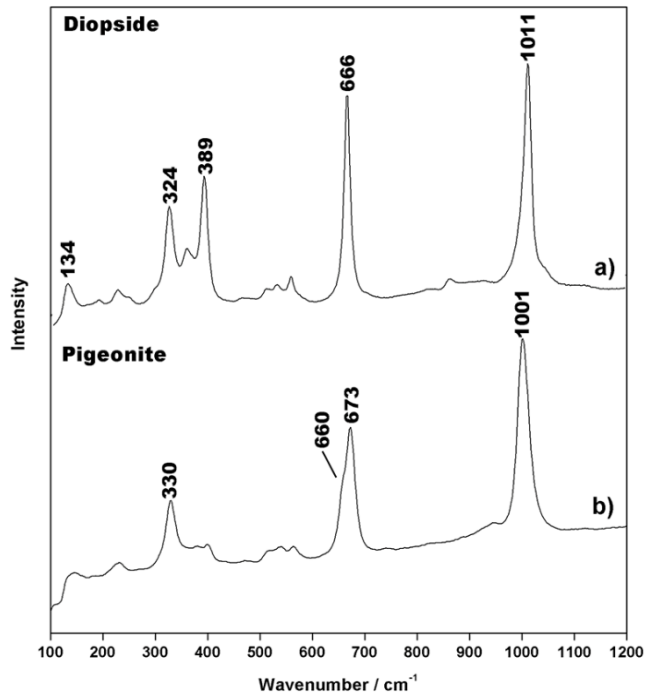

*Figure 08: Raman spectra of diopside (a) and pigeonite (b), collected from sample IS16-0013 by means of the laboratory InVia spectrometer.*

Compared to XRD and NIR results, Raman spectroscopy provided additional information regarding the composition of feldspar grains. Indeed, even though labradorite peaks at 177, 286, 477, 506, and 564 cm$^{-1}$ were detected in most of the analysed grains, some of the obtained spectra showed a slight displacement of the 177cm$^{-1}$ peak towards lower wavelengths (165 cm$^{-1}$) together with an additional signal of moderate intensity at 410 cm$^{-1}$ (Figure 09b). The observed vibrational features fit well with the ternary feldspar (K-Na-Ca) spectrum described in the work of Freeman et al. [40], which was obtained from a mineral sample with composition $Or_{26}Ab_{65}An_9$ (were Or, Ab and An are the end-members orthoclase (K), albite (Na) and anorthite (Ca) respectively).

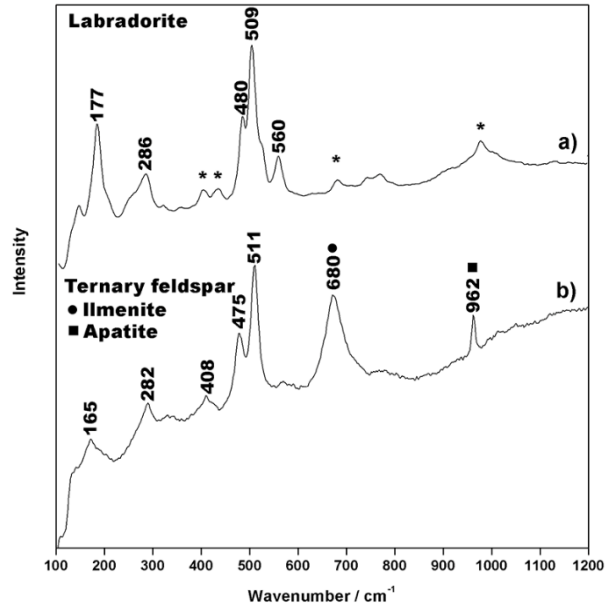

*Figure 09: Raman spectra of labradorite (a) and ternary feldspar (b), collected from sample IS16-0013 by means of the laboratory InVia spectrometer. In the lower spectrum, the main peaks of ilmenite and apatite are also present. * Pyroxene Raman signals.*

The only 3 grains of olivine detected in this sample showed their main doublet at lower wavelengths (818 and 847cm$^{-1}$) than the crystals from sample IS16-0002. Using the calibration curve proposed by T. Mouri et al. [32], a concentration of Mg (with respect to Fe) around 55% was calculated, indicating a solid solution of Fo$_{55}$Fa$_{45}$. As in the case of sample IS16-0001, ilmenite (Figure 09) was also detected as minor phase.

With regards to alteration products, the characteristic peaks of hematite were observed in at least 40% of the total spectra. In addition to hematite, apatite (Figure 09, phosphate Ca$_5$(PO$_4$)$_3$(F,Cl,OH), main peak at 962 cm$^{-1}$) was also found as secondary mineral product. Therefore, according to Raman data, the alteration of this sample seems to be stronger than IS16-0001 and IS16-0002.

**3.1.4 X-ray fluorescence of fine powders (M4-300+ Tornado)**

The mineralogical characterization of Icelandic basalts was complemented by their geochemical study. For this reason, 5 semi-quantitative analyses were carried out on fine-grained powdered samples by using the laboratory XRF system (spot diameter 1 mm). The results were then employed to calculate the average concentration values of each detected element as well as to establish the standard deviation of the measurements.

Under a qualitative point of view the composition of the samples is very similar. However, the semi-quantitative results represented in Figure 10 displays differences in the concentration of the detected elements. The major elements composing the three analogues are Si (from 30.7% ± 2.0% to 32.2% ± 1.8%), Fe (from 25.0% ± 2.1% to 31.9% ± 2.1%), Ca (from 20.8% ± 2.8 % to 25.1% ± 3.4%), Al (from 10.2% ± 1.4% to 11.3% ± 1.6%) and Mg (from 3.0% ± 0.3 to 7.3% ± 3.1%).

With regards to minor (concentration <1%) elements, titanium, manganese, chromium and phosphorus were detected in the three analogues. Furthermore, traces of nickel were detected in samples IS16-0001 and IS16-0002, vanadium in samples IS16-0002 and IS16-0013, while potassium was only found in sample IS16-0013.

Comparing the elemental composition of the three analogues and relating these data with the mineralogical information obtained from spectroscopic (NIR and Raman) and diffractometric (XRD) techniques, it can be deduced that the difference in the concentration of Ca in the samples is mainly related to the total content of pyroxenes and feldspars (the mineralogical groups that contain Ca in their chemical composition). In the case of pyroxenes, it is also necessary to take into account the concentration ratio between augite, diopside and pigeonite.

The content of Mg in sample IS16-0001 (7.2% ± 3.0%) is almost twice as high as measured in samples IS16-0002 (4.2% ± 0.7%) and IS16-0013 (3.0% ± 0.3%). This difference is mainly due to the high concentration of olivine ($(MgFe)_2SiO_4$) detected in the first analogue. IS16-0013 is the sample showing the highest concentration of iron, which fits with the frequent detection of hematite ($Fe_2O_3$) and to the Fo/Fa ratio of the analyzed olivine crystals. On the other hand, the titanium content in samples IS16-0002 and IS16-0013 is higher than that detected in IS16-0001. In this case, this difference can be attributed to the ilmenite crystals detected in both samples through NIR and Raman spectroscopy.

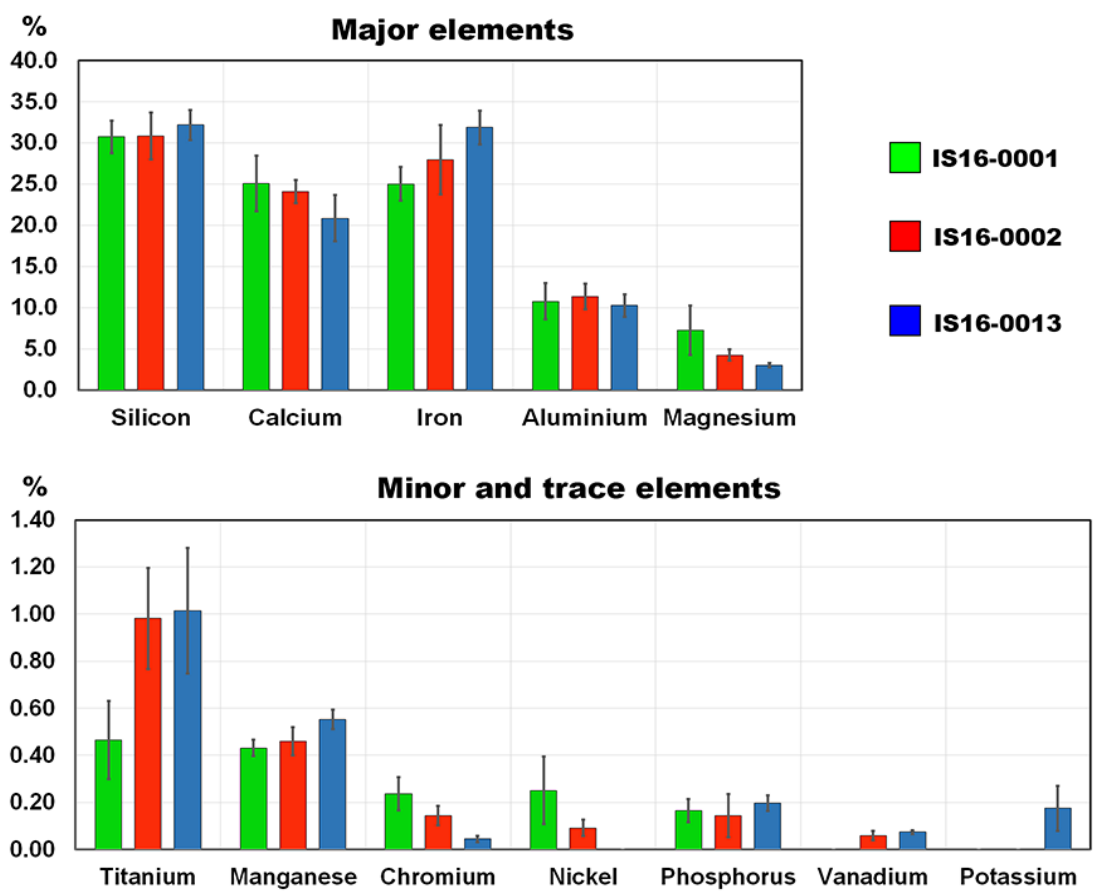

*Figure 10: Semi-quantitative elemental data obtained from the study of powdered samples by means of the XRF spectrometer.*

**3.2 Analysis with the flight-derived analytical instrument**

**3.2.1 Raman spectroscopy of coarse-grained powders (RLS ExoMars Simulator)**

As detailed in section 2.2.2, the RLS ExoMars Simulator was employed to evaluate the accuracy of the algorithms developed to perform automatic analysis on Mars. In addition, by comparing the obtained results with those provided by the InVia laboratory system, inferences regarding the scientific capabilities of the RLS on Mars can be extrapolated.

In a general perspective in must be highlighted that, unlike InVia results, RLS ExoMars Simulator analysis often detected multiple compounds in the same spectrum. This characteristic is due to the fact that the two instruments were coupled to objectives of different magnification (50x for the RLS ExoMars Simulator and 100x for the InVia), which has a direct effect on the diameter of the spot of analysis: 50 microns and 20 microns respectively. In addition, it must be highlighted that RLS ExoMars Simulator analyses were carried out by automatically moving the positioners in the x-axis at steps of 150 microns. For this reason, the excitation laser was frequently focused in areas of heterogeneous composition while, in the case of InVia measurements, the spot of analysis was manually focused on selected single crystals.

In detail, concerning the identification of major minerals from sample IS16-0001, RLS ExoMars Simulator results confirmed the detection of the major mineral groups identified through the InVia system, i.e. pyroxene, feldspar and olivine.

In the case of pyroxenes, besides the main characteristic peaks at 666 and 1010 cm$^{-1}$, through the detection of secondary peaks two compounds were identified: the peaks at 326, and 391 cm$^{-1}$ are characteristic of augite (Figure 11a), while the additional peak at 134 cm$^{-1}$ confirmed the detection of diopside (Figure 11b).

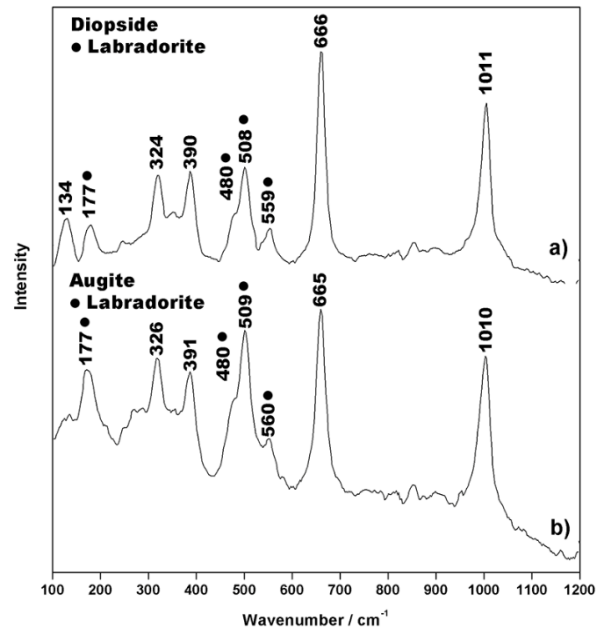

*Figure 11: Raman spectra of diopside (a) and augite (b), collected from sample IS16-0001 by means of the RLS ExoMars Simulator. In both spectra, the main peaks of labradorite are clearly observed.*

All feldspar spectra shown the same vibrational profile, characterized by intense peaks around 177, 480, 509 and 560 cm$^{-1}$ (Figure 11b). In spite of the intense Raman signals, the main peaks at 480 and 509 cm$^{-1}$ are not well resolved as those represented in Figures 5 and 9. This difference is mainly due to the difference between the spectra resolution of InVia (1 cm$^{-1}$) and RLS ExoMars Simulator systems (6-10 cm$^{-1}$). As explained above, the constant detection of the strong peak at 177 cm$^{-1}$ evidences the identification of labradorite as one of the major mineralogical phases of this sample.

Unlike InVia analysis, the characteristic spectra of anorthite was not detected by the RLS ExoMars Simulator. This datum highlights that anorthite represents a minor or trace fraction of sample mineralogy.

In a few spectra, a weak peak at 1086cm$^{-1}$ was detected, confirming the presence of calcite. Similarly, a further peak around 142 cm$^{-1}$ was also observed. This signal could be assigned to anatase (TIO$_2$) which, together with ilmenite identified by NIR, fits with the detection of Ti as trace element.

In the case of olivine, the RLS ExoMars Simulator was perfectly capable of identifying the displacement of the main double peak. In terms of peaks position, RLS Simulator data fitted with those obtained from Raman InVia, measuring a shifting between 817 and 825 cm$^{-1}$ ($\kappa_2$ peak) and 849 and 855 cm$^{-1}$ ($\kappa_1$ peak) respectively. These data strengthen the hypothesis according to which the composition of olivine crystals varies between Fo$_{86}$Fa$_{14}$ and Fo$_{64}$Fa$_{36}$. Concerning the detection of alteration products, hematite peaks of relative low intensity were detected in some of the collected spectra.

The RLS ExoMars Simulator analysis of sample IS16-0002 confirmed the detection of pyroxene, feldspar and olivine as major mineralogical phases of the analogue. With respect to pyroxenes and feldspars, the data obtained from this sample perfectly coincide with those of sample IS16-0001, confirming the detection of augite and diopside (pyroxene) as well the identification of labradorite (feldspar).

Compared to the InVIa results, the RLS ExoMars Simulator provided additional information about olivine composition. Even though most of the spectra (around 25) confirmed the detection of olivine grains with the two main peaks located at 822 and 854 cm$^{-1}$ (Figure 12a), the main peaks shifting towards lower wavelengths (down to 818 and 849 cm$^{-1}$, see Figure 12b) was detected in some of the analysed spots, indicating that a minor fraction of olivine crystals had a higher concentration of iron ions.

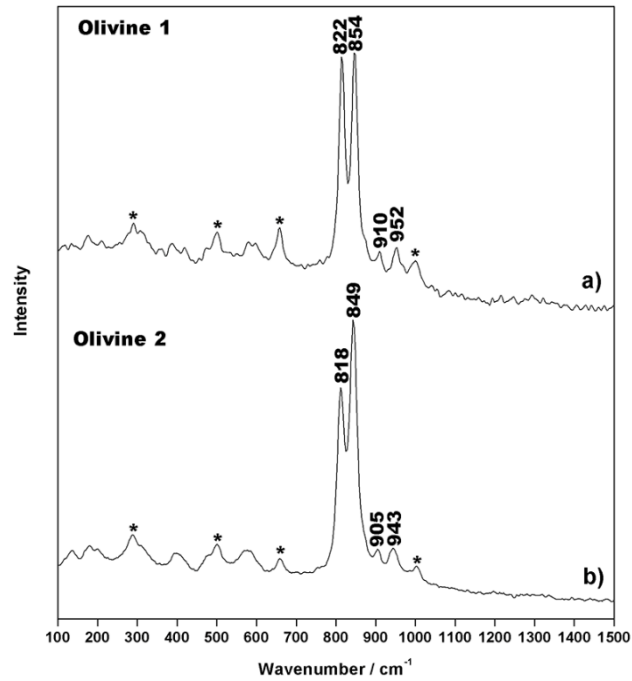

*Figure 12: Raman spectra of different olivine grains from sample IS16-0002, analyzed by means of the RLS ExoMars Simulator. The displacement of the main Raman peaks is related to the Fo/Fa ratio of the solid solution. * Raman signals proceeding from additional compounds.*

Concerning sample IS16-0013 results, pyroxene and feldspar were detected in most of the RLS ExoMars Simulator analysis. With regards to pyroxene, both augite and diopside were identified, while pigeonite (detected in a few spots by Raman InVia spectrometer) was not observed. Similarly, all feldspar spectra fitted with the pattern of labradorite while the characteristic vibrational features of ternary feldspar (detected by Raman InVia) were not detected. In a limited number of spectra, additional peaks of medium and low intensity were found at 822 and 854 cm$^{-1}$, confirming the presence of olivine. In this case, although the quality of the spectra was poorer than those obtained for samples IS16-0001 and IS16-0002, the position of the double peak coincided with the values obtained by InVia system, confirming the presence of olivine $Fo_{83}Fa_{17}$.

To conclude, this analogue is distinguished from the previous ones by almost constant detection of alteration products in form of hematite. Furthermore, as shown in Figure 13, the detection of apatite was also confirmed.

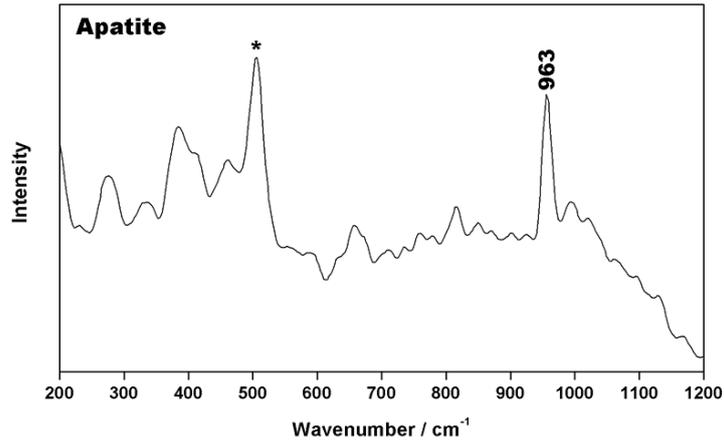

*Figure 13: Raman spectra of apatite, collected from sample IS16-0013 by means of the RLS ExoMars Simulator. * Raman signals proceeding from additional compounds.*

The general results obtained from each sample by both commercial and flight-derived analytical systems are summarized in Table 01.

*Table 01: mineralogical results comparison (the values included between parentheses represent the semi-quantitative estimation obtained by each technique).*

| Detected compounds | | IS16-0001 Commercial instruments | | | IS16-0001 RLS ExoMars Simulator | IS16-0002 Commercial instruments | | | IS16-0002 RLS ExoMars Simulator | IS16-0013 Commercial instruments | | | IS16-0013 RLS ExoMars Simulator |
|---|---|---|---|---|---|---|---|---|---|---|---|---|---|
| | | XRD | NIR | Raman | | XRD | NIR | Raman | | XRD | NIR | Raman | |
| Pyroxene | Diopside | | X (44%) | X | X | | X (35%) | X | X | | | X | X |
| | Augite | X (35%) | X (9%) | X | X | X (48%) | X (33%) | X | X | X (50%) | X (78%) | X | X |
| | Pigeonite | | | | | | | | | | X (6%) | X | |
| Feldspar | Plagioclase | X (40%) | X (44%) | X | X | X (40%) | X (30%) | X | | X (40%) | X (15%) | X | X |
| | Ternary | | | | | | | | | | | X | |
| | Anorthite | | | | X | | | | | | | | |
| Olivine | Forsterite | X (25%) | X (<5%) | X | X | X (11%) | X (<5%) | X | X | X (3%) | | | X |
| Ti/Fe-oxides | Ilmenite | X (<5%) | X (<1%) | X | | | X (2%) | X | | | X (<1%) | | |
| | Anatase | | | | X | | | | | | | | |
| | Hematite | X (<5%) | | X | | | | X | | | | X | X |
| Carbonates | Ankerite | X (<5%) | | | | X (<5%) | | | | X (<5%) | | | |
| | Calcite | X (<5%) | | x | X | X (<5%) | | | | X (<5%) | | | |
| Others | Carbon | | | X | | | | X | | | | | |
| | Apatite | | | | | | | | | | | X | X |

## 4 Conclusions

The analysis presented in this study enables the detailed characterization of the selected terrestrial analogues, providing the whole set of data necessary to feed the PTAL database.

Concerning the results obtained from commercial analytical systems, it must be underlined that the combination of NIR and Raman data fitted well with diffractometric results. Specifically, the NIR technique ensured an accurate overview of the mineralogy diversity of the sample by providing semi-quantitative results, while the Raman technique was able to detect minor compounds that were not identified by NIR and XRD analysis. Considering that the Analytical Laboratory Drawer of the ExoMars rover will provide the possibility of performing combined or cooperative science between the MicrOmega and RLS spectroscopic system, to prove the complementarity of NIR and Raman data in the mineralogical determination of Martian analogues is a promising result in the perspective of the forthcoming ExoMars/ESA mission.

With regards to the Raman characterization of the samples, RLS ExoMars Simulator results fit with those provided by the state-of-the-art laboratory system. Considering that the RLS ExoMars Simulator has been used by simulating the analytical routine that the RLS will follow on Mars, this work confirms that a full picture of the mineralogical heterogeneity of powdered samples can be obtained through the automatic acquisition of 40 spectra.

Among the obtained results, it must be underlined that the RLS ExoMars Simulator was able to clearly detect the whole set of spectroscopic features (main and secondary peaks) of all main phases (a detail of critical importance in the identification of the correct phase within a mineral group). More important, the RLS ExoMars Simulator was also capable of detecting the shift of the characteristic double peak (820-850 $cm^{-1}$) of olivine. Considering that this phenomenon is linked to the Fe/Mg ratio within its crystalline structure, this work suggests that the ExoMars/RLS has the potential to provide detailed information about the elemental composition of olivine on Mars.

From the promising results obtained in this study, the RLS science group intends to deepen the study of the double peak shifting. The purpose is to formulate an empiric equation that, taking into account the analytical features of the RLS, will allow to precisely calculate the Fo/Fa ratio from the olivine spectra that will be collected on Mars.

## Acknowledgements

This project is financed through the European Research Council in the H2020- COMPET-2015 programme (grant 687302) and the Ministry of Economy and Competitiveness (MINECO, grants ESP2014-56138-C3-2-R and ESP2017-87690-C3-1-R).